# Room temperature magnetic order on zigzag edges of narrow graphene nanoribbons

*Gábor Zsolt Magda[1], Xiaozhan Jin[2], Imre Hagymási[3,4], Péter Vancsó[1], Zoltán Osváth[1], Péter Nemes-Incze[1], Chanyong Hwang[2], László P. Biró[1] and Levente Tapasztó[1]\**

1. Research Centre for Natural Sciences, Institute of Technical Physics and Materials Science, Nanotechnology Department, 2D Nanoelectronics „Lendület" Research Group, Budapest, Hungary
2. Korea Research Institute of Standards and Science, Center for Nanometrology, Daejeon, South Korea
3. Wigner Research Centre for Physics, Institute of Solid State Physics and Optics, Strongly Correlated Systems "Lendület" Research Group, Budapest, Hungary
4. University of Szeged, Department of Theoretical Physics, Szeged, Hungary

*e-mail: tapaszto@mfa.kfki.hu

**Magnetic order emerging in otherwise non-magnetic materials as carbon is a paradigmatic example of a novel type of *s-p* electron magnetism predicted to be of exceptional high-temperature stability[1]. It has been demonstrated that atomic scale structural defects of graphene can host unpaired spins[2,3]. However, it is still unclear under which conditions long-range magnetic order can emerge from such defect-bound magnetic moments. Here we propose that in contrast to random defect distributions, atomic scale engineering of graphene edges with specific crystallographic orientation – comprising edge atoms only from one sub-lattice of the bipartite graphene lattice – can give rise to a robust magnetic order. We employ a nanofabrication technique[4] based on Scanning Tunneling Microscopy to define graphene nanoribbons with nanometer precision and well-defined crystallographic edge orientations. While armchair ribbons display quantum confinement gap, zigzag ribbons narrower than 7 nm reveal a bandgap of about 0.2 - 0.3 eV, which can be identified as a signature of**

**interaction induced spin ordering along their edges. Moreover, a semiconductor to metal transition is revealed upon increasing the ribbon width, indicating the switching of the magnetic coupling between opposite ribbon edges from antiferromagnetic to ferromagnetic configuration. We found that the magnetic order on graphene edges of controlled zigzag orientation can be stable even at room temperature, raising hope for graphene-based spintronic devices operating under ambient conditions**.

The intrinsic magnetism of graphite has a long and controversial history[1]. The origin of the measured magnetic signal is generally attributed to atomic scale structural defects locally breaking the sub-lattice balance of the bipartite hexagonal lattice[5,6]. However, the unambiguous identification of the structural sources of the measured magnetic signal has proven challenging as they are buried inside the bulk of the material. The isolation of single graphene layers[7] opens new prospects in this direction[8,9] as their atomic structure is fully accessible for imaging and controlled modification. In particular, graphene edges of specific (zigzag) crystallographic orientation comprising carbon atoms from only one sub-lattice of the bipartite hexagonal lattice are predicted to host magnetic order[10], in striking contrast to armchair edges incorporating an equal number of carbon atoms from both sub-lattices.

The strong influence of edge orientation on the electronic structure of graphene nanoribbons had been predicted[11] long before graphene was isolated[7]. However, a systematic experimental verification of this fundamental prediction is still lacking, due to the very limited control over the edge orientation of the investigated graphene ribbons. Opening of a gap inversely proportional to the ribbon width has been experimentally demonstrated by electrical transport measurements[12,13]; however, no evidence on the effect of edge orientation has been revealed. This is mainly attributed to the random edge orientations and the presence of a considerable, nanometer scale edge roughness, suppressing orientation effects[14]. STM study of irregularly shaped graphene ribbons revealed that structures with a higher ratio of zigzag edges display smaller bandgaps as compared to those where armchair edges are in majority[15]. This is clearly indicative of edge specific physics; however, the lack of orientation control did not allow a more systematic insight. From the theory side, there is a broad consensus that graphene nanoribbons with armchair edges are semiconductors, due to the quantum confinement of the charge carriers, while zigzag ribbons host peculiar flat-band edge states[10,16]. In the non-interacting picture, these edge states localized near the Fermi level render all zigzag graphene nanoribbons (z-GNRs) metallic. However, such 1D metallic edge states with a high local density of states at the Fermi level become unstable upon electron-electron interactions. To lower the energy of the system, it is favorable to split the flat band (open a gap) by ordering the spins along the two ribbon edges with antiferromagnetic coupling between opposite edges. Consequently, the emergence of magnetic order is closely linked to altering the electronic structure of the ribbons, through opening a bandgap in the otherwise metallic zigzag nanoribbons[10,16,17]. This enables us to detect the signature of edge magnetism on individual graphene nanostructures by investigating their electronic structure, as directly measuring magnetic signals would require a macroscopic amount of such ribbons. The edge-state magnetism and the associated bandgap opening in zigzag ribbons is consistently predicted by various theoretical models, including first principles DFT[16], mean-field theory based Hubbard[18] and Quantum Monte Carlo calculations[19], indicating that the edge-magnetism is a robust property of z-GNRs, not sensitive to the specific details of the models. However, the stability of the magnetic order on real graphene edges and experimental conditions is strongly debated. Experimental indications that edge magnetism can indeed occur at low temperatures (7K) have been provided by tunneling spectroscopy measurements on graphene ribbons obtained by unzipping carbon nanotubes[20]. However, the random orientation of the edges and the influence of a possible strong edge-substrate hybridization[21] did not allow full access to the nature of edge-magnetism in graphene. Though the magnetic order is expected to persist to some extent on zigzag segments of randomly oriented graphene edges, the mixing of different edge types are expected to substantially weaken the effect[19,22]. Therefore, the lack of experimental control over the edge orientation seems one of the main reasons that the magnetic graphene edge states consistently predicted by various theoretical models remained experimentally so elusive.

To realize graphene nanoribbons with precisely controlled crystallographic edge orientations, we employ a nanofabrication technique based on scanning tunneling microscopy[4]. By acquiring atomic resolution images, the crystallographic directions of the graphene lattice can be identified and matched with the desired cutting directions. The direct cutting of graphene is done by the beam-induced chemical etching activated by the sub-nanometer wide channel of tunneling electrons, locally breaking the carbon-carbon bonds underneath the atomically sharp STM tip apex. More details on the nanolithographic process are given in the Methods section.

Using the STM nanolithographic technique, we have defined nanoribbons into chemical vapor deposition grown graphene sheets on gold substrates with large Au (111) terraces. Au (111) was predicted to be one of the few substrates preserving edge magnetism in supported graphene ribbons[21]. Nanoribbons with pre-defined armchair or zigzag edge orientation and widths ranging down to 3 nm have been defined (Fig.1). The as-fabricated ribbons display regular edges of sub-nanometer roughness. The bright protrusions near the edges in Fig. 1 are due to a known imaging instability[23], rather than edge defects or impurities. Upon optimizing the STM imaging the edges defined by STM lithography were found to be straight and close to atomically smooth, free of detectable impurities, reconstructions or curvature (Extended data: Fig.1).

Tunneling spectra acquired on armchair ribbons displayed semiconducting I-V characteristics, in contrast to spectra taken on graphene outside the ribbons (see Extended data: Fig. 2). The measured bandgap as a function of the corresponding ribbon width is plotted in Fig 2a. A clear inverse proportionality is revealed, in excellent quantitative agreement with our calculations based on the Hubbard model (Eq. 1) considering hydrogen saturated and relaxed edges, as well as first principles DFT calculations[16]. The theoretical data points correspond to three classes of armchair ribbons those with $3n$, $3n+1$ and $3n+2$ ($n = 1,2,3,…$) rows of carbon dimers across their width. The measurement error and the eventually present atomic scale edge irregularities do not allow us to clearly distinguish between ribbons belonging to the 3n and 3n+1 class; however, armchair ribbon segments with no detectable gap have been observed in a few cases, which most probably belong to the 3n+2 class. Consequently, our measurements systematically verify the predicted mechanism of bandgap opening due to the lateral confinement of the charge carriers in ribbons of precisely armchair edge orientation.

A strikingly different behavior has been revealed in zigzag graphene nanoribbons (Fig.2b). For z-GNRS narrower than 7 nm a fairly large bandgap of about 200 - 300 meV has been observed. As discussed above, the broadly predicted origin of bandgap opening in zigzag ribbons are electron-electron interactions and so far the only predicted many-body ground state implies the magnetization of the edges[16,18,19].

However, in contrast to first principles theoretical predictions[16] the measured gap suddenly vanishes for zigzag ribbons wider than 8 nm. We attribute this discrepancy to the fact that these calculations have been performed at zero temperature and without doping, while the experimental data has been acquired at room temperature and finite doping ($\Delta E_F \sim +50 – +100$ meV, see Extended data: Fig. 2c). Without considering the effects of temperature and doping, the ground state was found to be always semiconducting (antiferromagnetic) and no steep transition to a metallic state is expected. To understand the origin of the experimentally observed sharp semiconductor-metal transition, we have performed calculations based on the mean-field approximation of the Hubbard Hamiltonian that accounts for both finite temperature and doping:

$$\mathcal{H} = -\sum_{\langle i,j \rangle} t_{ij}\hat{c}_{i\sigma}^{\dagger}\hat{c}_{j\sigma} + U\sum_{j}\hat{n}_{j\uparrow}\hat{n}_{j\downarrow} - \mu\hat{N}, \qquad (1)$$

where the first term is the nearest-neighbor tight-binding Hamiltonian, while the second term stands for the onsite Coulomb repulsion. The effects of temperature and doping have been included using the grand canonical ensemble (third term), where $\mu$ is the chemical potential, and $\hat{N}$ the particle number operator. The electron density and the chemical potential were determined self-consistently. (For more details on calculations see the Methods section).

The main result of our extended theoretical model is that it reproduces the steep semiconductor-metal transition, quantitatively accounting for both the critical ribbon width (w ~ 7 nm) at which the transition occurs and the magnitude of the measured bandgaps, using the strength of the on-site repulsion ($U$) as the single unknown parameter (Fig.2b). From calculating the spin density distribution for both semiconducting (w < 7 nm) and metallic (w > 8 nm) ribbons the origin of the observed semiconductor→metal transition can be identified as a transition from an antiferromagnetic (semiconducting) state, where magnetic moments on opposite ribbon edges are aligned antiparallel, to a ferromagnetic (metallic) state, with parallel spin alignment on opposite edges (Fig.3). The metallic nature of z-GNRs with ferromagnetic inter-edge coupling has already been predicted [18] together with a complex magnetic phase diagram as a function of doping. Our results are consistent with these findings in the T → 0 K limit. Here, we note that the observation of the gap opening alone in z-GNRs would not unambiguously indicate edge-magnetism, as zigzag edge configurations (e.g. $z_{211}$) leading to semiconducting but non-magnetic z-GNRs have been predicted[24]. However, only magnetic zigzag edges can account for both gap opening and the observed semiconductor-metal transition. This, together with the quantitative agreement between our calculations and experiments provide indirect but compelling evidence that magnetic order emerges on zigzag graphene edges of precisely engineered crystallographic orientation. Moreover, it is truly remarkable that the signature of edge magnetism can be experimentally detected at room temperature. This is nevertheless in agreement with the expected high-temperature stability of *s-p* electron magnetism[25,26], as well as experiments reporting room temperature magnetism in defective graphitic samples[5,27,28,29].

Since the only free parameter used in the calculations for zigzag ribbons is the magnitude of the on-site repulsion ($U$), the best quantitative agreement with our experimental findings provides an experimental estimate for the strength of the electron-electron interaction in graphene of $U = 3.24$ eV. Theoretical works so far have predicted the magnitude of $U$ ranging from 2 eV to 6 eV[30]. The experimentally estimated value of $U$ is even more reliable as a single $U$ value can reproduce both the magnitude of the gaps and the semiconductor-metal transition ribbon width.

The fact that the experimental results can be quantitatively interpreted by theoretical calculations on perfect zigzag (armchair) edges confirms the experimentally found high edge-quality. However, deviations from the ideal edge structure can be present on the atomic scale. To estimate the effect of atomic scale edge irregularities, we have performed calculations on a model system comprising a high density of such defects (Extended data: Fig.3). The results indicate that even in the presence of a high defect density, for overall zigzag oriented edges the qualitative picture holds but the strength of the effect (gap size, spin polarization) decreases. In order to fit the experimental data, the value of electron-electron interactions estimated based on the ideal zigzag edge has to be increased from $U = 3.24$ eV to $U = 4.32$ eV.

Our findings demonstrate that graphene nanoribbons display strongly edge orientation specific behavior and engineering the crystallographic orientation of graphene edges allows us an unprecedented control over both electronic and magnetic properties of graphene nanostructures, opening the way towards the realization of electronic and spintronic devices based on robust quantum mechanical effects, enabling their room temperature operation.

## Acknowledgements

The experimental work has been conducted within the framework of the Korea Hungary Joint Laboratory for Nanosciences through Korean Research Council of Fundamental science and technology and the "Lendület" program of the Hungarian Academy of Sciences. L.T. acknowledges OTKA grant K 108753 and the Bolyai Fellowship. L.P.B. acknowledges OTKA grant K101599. C.H. is partially supported by Nano·Material Technology Development Program through the National Research Foundation of Korea (NRF) funded by the Ministry of Science, ICT and Future Planning (2012M3A7B4049888). I.H. was supported by European Union and the State of Hungary, co-financed by the European Social Fund in the framework of TÁMOP-4.2.4.A/ 2-11/1-2012-0001 'National Excellence Program and OTKA grant K100908. I.H. acknowledges discussions with K. Itai. L.T and P.V. acknowledge discussions with Y.-S. Kim..

## Author contributions

L.T. and C.H conceived and designed the experiments. G.Z.M. performed the lithography and STM experiments. I.H. and P.V. provided the theoretical results. X.J. and C.H. performed the graphene growth experiments. Z.O. and P.N.I. carried out preliminary experiments. L.T., P.V., I.H., G.Z.M. and L.P.B. analyzed the data. L.T. wrote the paper. All the authors discussed the results and commented on the manuscript.

## Author Information

Correspondence and requests for materials should be addressed to L.T. (tapaszto@mfa.kfki.hu)

## Methods

The graphene samples have been grown by Chemical Vapor Deposition on Cu foil, and transferred to a stripped gold substrate with large, atomically flat Au (111) terraces. The lithographic process and tunneling microscopy/spectroscopy investigations have been performed using a Nanoscope E Scanning Tunneling Microscope operating under ambient conditions. First, atomic resolution STM images of the graphene lattice were acquired, ($U_{bias}$ = 5 - 50 mV and $I_{tunnel}$ = 1 - 2 nA) to precisely identify the zigzag/armchair directions of the graphene lattice. For patterning, a bias voltage of 2.0 – 2.3 V (tip negative) is applied between the tip (Pt-Ir, 90% - 10%) and the sample while the tip is slowly (1-5 nm/s) moved along the desired cutting direction. The humidity of the cutting atmosphere (70-75%) can be precisely controlled inside an atmospheric hood. Before the tunneling spectroscopy measurements on ribbons, tunneling spectra have been acquired on gold to ensure about their linearity and confirm a good quality, contamination free tip. We have acquired tunneling spectra inside the ribbons to confirm the presence/absence of a bandgap. This method has proven more efficient and reproducible under ambient than measuring strongly edge-localized peaks in the dI/dV spectra. We define the gap of the ribbon as the width of the plateau around zero bias in tunneling spectra (Extended data: Fig. 2b). These plateaus are not entirely flat, but have a shallow slope, which we attribute to the presence of gold substrate underneath the atomically thin graphene nanostructures. The uncertainty (error bars) in the measured gap values mainly come from the thermal broadening and the influence of the substrate, while for ribbon widths it is due to the STM tip convolution effects. As control experiments, both before and after measuring the I-V characteristics of a ribbon, tunneling spectra have been acquired on the graphene sheet outside the ribbons. The (non-linear) spectra measured on the ribbons were only taken as reliable if both before and after their measurement, the spectra of the unpatterned graphene displayed the expected closely linear I-V characteristics. The tunneling I-V characteristics of graphene and metallic graphene nanoribbons on gold substrate are closely linear due to the contribution of a high local density of state of the Au substrate. Within (along) the ribbons the measured spectra were fairly homogeneous (Extended data: Fig.4), apart from a few exceptions, which we attribute to locally present edge defects or impurities.

To interpret the experimental results we considered the following Hamiltonian:

$$\mathcal{H} = -\sum_{\langle i,j \rangle} t_{ij} \hat{c}_{i\sigma}^{\dagger} \hat{c}_{j\sigma} + U \sum_{j} \hat{n}_{j\uparrow} \hat{n}_{j\downarrow} - \mu \widehat{N}$$

which was introduced in the main text. We applied the

$$\hat{n}_{i\sigma} = \hat{n}_{i\sigma} - \langle \hat{n}_{i\sigma} \rangle + \langle \hat{n}_{i\sigma} \rangle$$

identity and neglected the term which contains the fluctuations. Quantum Monte Carlo (QMC) simulations including fluctuations, found the agreement between mean field theory (MFT) and QMC to be remarkably accurate for moderate Coulomb interactions, justifying the application of MFT for the description of realistic ribbon geometries[31]. We arrived at a single particle problem, which can be diagonalized by a generalized Bogoliubov transformation in *k* space:

$$\mathcal{H}_{MF} = \sum_{k,\sigma,n} (\varepsilon_{nk\sigma} - \mu) \hat{C}_{n,k,\sigma}^{\dagger} \hat{C}_{n,k,\sigma} - U \sum_{j} \langle \hat{n}_{j\uparrow} \rangle \langle \hat{n}_{j\downarrow} \rangle,$$

where $\hat{C}_{n,k,\sigma}^{\dagger} (\hat{C}_{n,k,\sigma})$ are the transformed operators that destroy (create) a particle with wavenumber *k* with spin $\sigma$ in band *n*. The energy bands are given by $\varepsilon_{nk\sigma}$, which depends on the yet unknown electron densities and chemical potential. The densities are calculated self-consistently, while the chemical potential is determined by the conservation of the number of particles. The self-consistent iteration was stopped when the difference between the electron densities was smaller than $10^{-12}$. It is known from the theory of Hubbard-model, that the half-filled case shows antiferromagnetic correlations for U>0, and at a certain $U_c$ a Mott-transition can take place. However, in the doped case both antiferromagnetic and ferromagnetic ground states can be obtained. The solution which has the lowest free-energy was accepted as the ground-state.

In order to provide accurate band gap values of the graphene nanoribbons, first we calculated the relaxed edge geometries with hydrogen passivation of narrow (2nm) zigzag and armchair ribbons. We used density functional theory based molecular dynamics calculations within the framework of local density approximation (LDA) using the VASP[32,33] simulation package. With the help of the relaxed edge coordinates of the atoms we were able to parameterize the tight binding hopping elements in our Hubbard-model. These hopping elements were applied to ribbons of various widths.

In the VASP calculations projector augmented wave (PAW) pseudo-potentials[34,35] were used and the kinetic energy cut-off for the plane wave expansion was 400 eV. In all ribbon geometries the atomic positions were relaxed using the conjugate–gradient method until the forces of the atoms were reduced to 0.02 eV/Å. We used a rectangular super-cell with 40 Å in the *x* and 2.46 Å (4.26 Å) in the *y* direction for z-GNR (a-GNR). The Brillouin zone was sampled using 2 *k*-points along the *x* axis and approximately 8/*dy* *k*-points along the *y* axis (*dy* in nm). Vacuum layers of 20 Å in the ribbon plane and 24.6 Å in the normal direction were applied in order to avoid interactions between nanoribbons in different unit cells.

From the relaxed carbon-carbon distances, we computed the hopping amplitudes using the parameterization.[36]

$$\tau(r_{ij}) = \left(\frac{r_{ij}}{a_0}\right)^{-\alpha 2} \exp[-\alpha 3(r_{ij}{}^{\alpha 4} - a_0{}^{\alpha 4})]$$

where $r_{ij}$ is the distance between the atoms *i* and *j*, $a_0$=1.42 Å (the carbon-carbon distance in the bulk), $\alpha_2 = 1.2785$, $\alpha_3 = 0.1383$ and $\alpha_4 = 3.4490$.

To estimate the maximum possible quantitative effect of edge irregularities on the magnetism, we have performed the calculations presented here on a model system with edges of overall zigzag orientation, but containing the maximum amount of atomic scale edge-defects that does not completely destroy edge magnetism (Extended data Fig.3). We found that the qualitative picture of edge magnetism including the semiconducting (AF) to metallic (FM) transition holds true for the defective zigzag edges, but the effect is substantially weakened as compared to perfect edges. Particularly, the calculated gap values are reduced. Therefore, the best fit to the experimental data occurs for a higher parameter value of the onsite repulsion parameter of $U = 4.32$ eV. Also the width dependence of the gap becomes weaker, but this is still in good agreement with the measurements (Extended data Fig.3b) for the increased $U$ parameter value.

# Figures

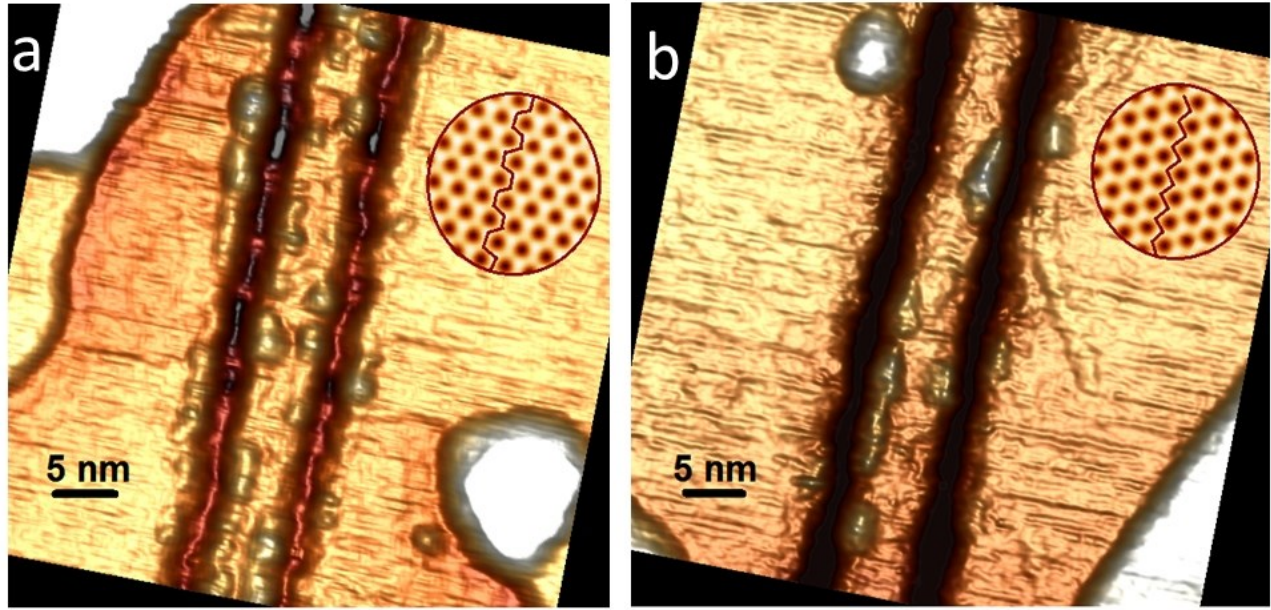

Figure 1

*Fig.1. Fabrication of graphene nanoribbons with precisely defined crystallographic edge orientations*. a) STM image (500 mV, 0.8 nA) of a 5 nm wide graphene nanoribbon with armchair edge orientation, and b) a 6.5 nm wide ribbon with edges of precisely zigzag orientation (300 mV, 2nA) patterned by scanning tunneling lithography in a graphene sheet deposited on Au(111) substrate. The circular insets show the atomic resolution STM images confirming the crystallographic directions of the edges. The atomic resolution images in the insets have been Fourier filtered for clarity. The protrusions on the otherwise highly regular edges are imaging artefacts.

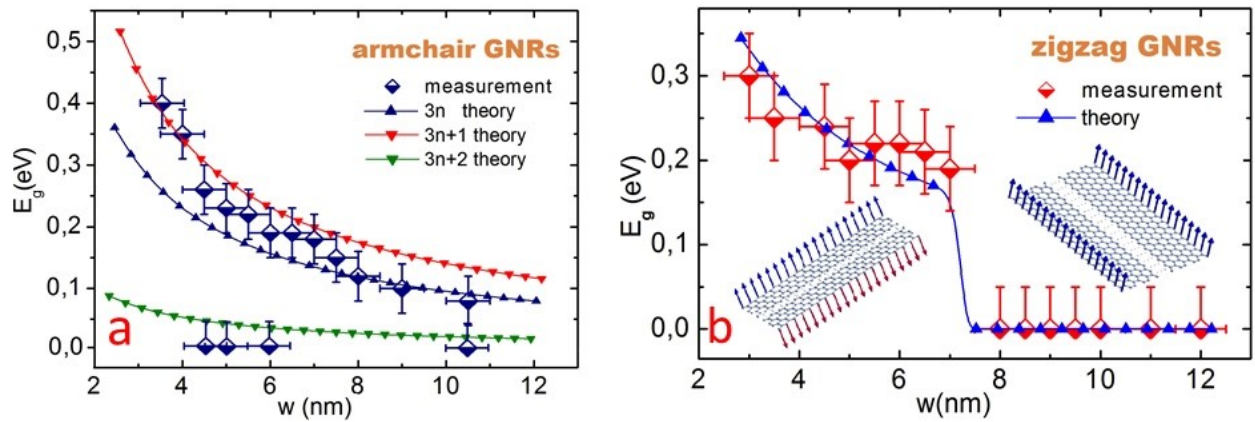

**Figure 2**

*Fig.2. Edge specific electronic and magnetic properties of graphene nanoribbons. The bandgap measured by tunneling spectroscopy as a function of ribbon width in armchair (a) and zigzag (b) ribbons. Armchair ribbons display a quantum confinement gap inversely proportional to their width. In zigzag ribbons the band structure is governed by the emerging edge magnetism and a sharp semiconductor (antiferromagnetic) to metal (ferromagnetic) transition is revealed. Theoretical data points have been calculated using the mean field Hubbard model (continuous lines are only guides to the eye). Error bars of the measured gap values originate from thermal broadening and substrate effects, while for ribbon width (horizontal) from tip convolution effects.*

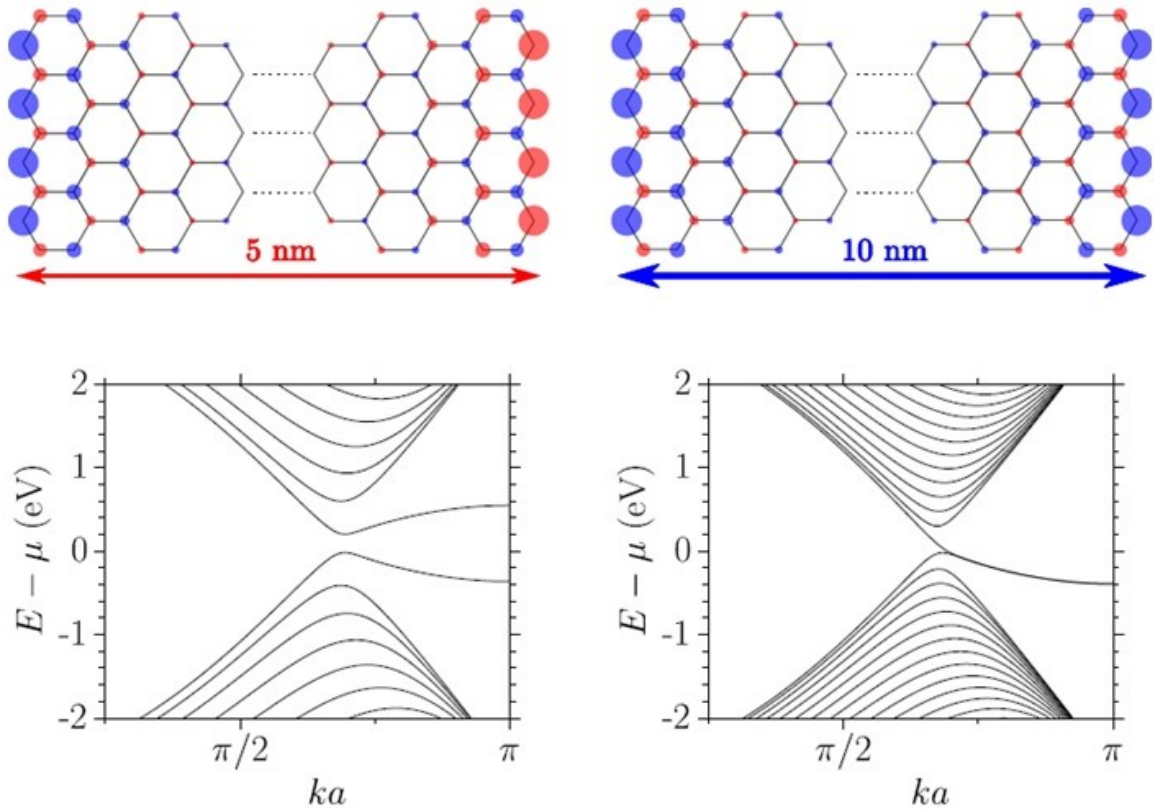

*Fig.3. **Correlating electronic and magnetic properties of zigzag graphene nanoribbons**. Spin density distribution (↑ - blue, ↓ - red) in a 5 and 10 nm wide zigzag graphene nanoribbon calculated in the mean field Hubbard model for T = 300 K and $\Delta E_F \sim 100$ meV. Lower panels display the corresponding band strcuture, clearly indicating that narrow zigzag ribbons are antiferromagnetic semicondcutors, while wider (w > 8nm) zigzag ribbons display a ferromagnetic inter-edge coupling and no bandgap.*

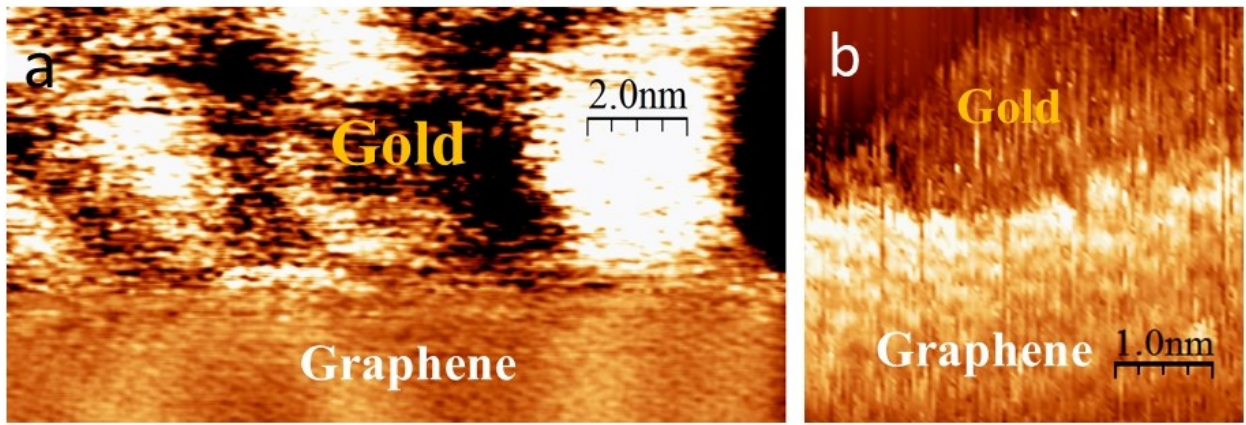

**Extended data: Figure 1**

**Extended data figure 1: The structure of graphene edges defined by STM lithography.** *a) STM image (5mV, 2nA) of 15 nm long edge segments cut by STM lithography, revealing close to atomically smooth edges (<5 A edge roughness) free of detectable reconstructions, contaminations or curvature. b) The increased local density of states on zigzag edges observed under specific imaging conditions (200 mV, 2nA) can be attributed to the presence of edge states that rules out the possibility of $sp^3$ type edge terminations (e.g., di-hydrogenated edges), as no edge states are expected to occur for such edge configurations.*

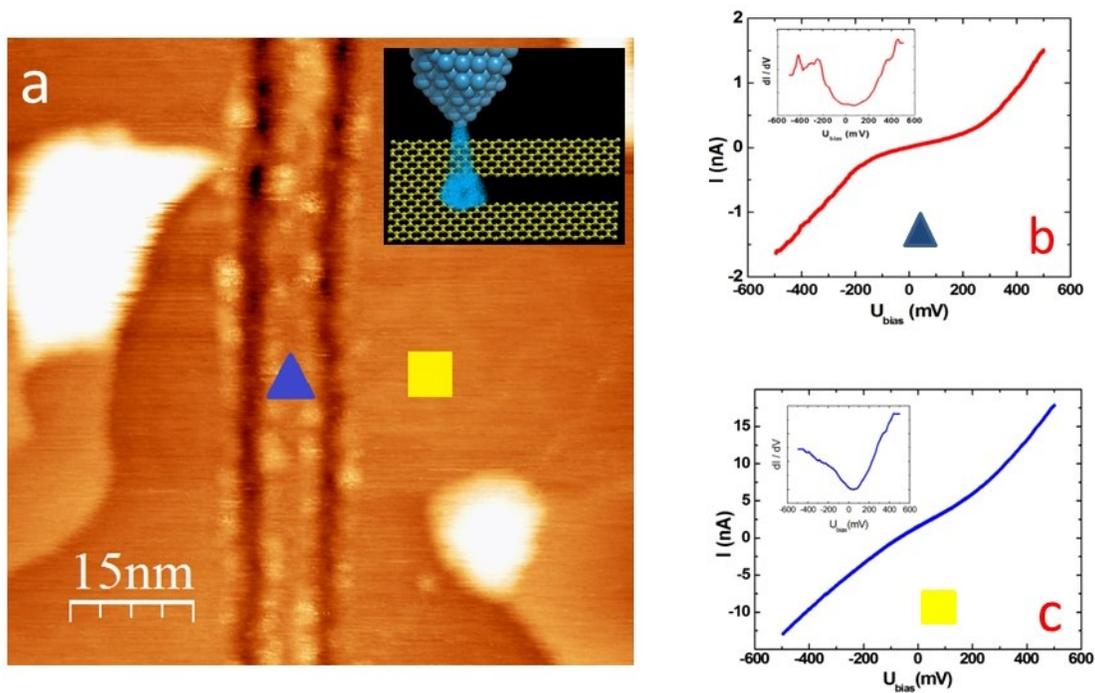

**Extended data: Figure 2**

**Extended data figure 2: Scanning tunneling spectroscopy of graphene nanoribbons on Au (111).** *Tunneling current-voltage (I-V) characteristics acquired on a 5 nm wide armchair ribbon (a) displaying nonlinear I-V spectra corresponding to a gap of about 250 meV (b). Outside the ribbon a closely linear characteristic of the unpatterned graphene is revealed (c). The insets show the schematics of STM lithography (a), and the differential tunneling conductance (dI/dV) obtained as numerical derivatives of the measured I-V curves (b,c). The ~ 70 mV shift of the Dirac point (curve minimum) from the Fermi level (zero bias) observed on graphene (c-inset) is due to the doping from the Au(111) substrate and the ambient atmosphere.*

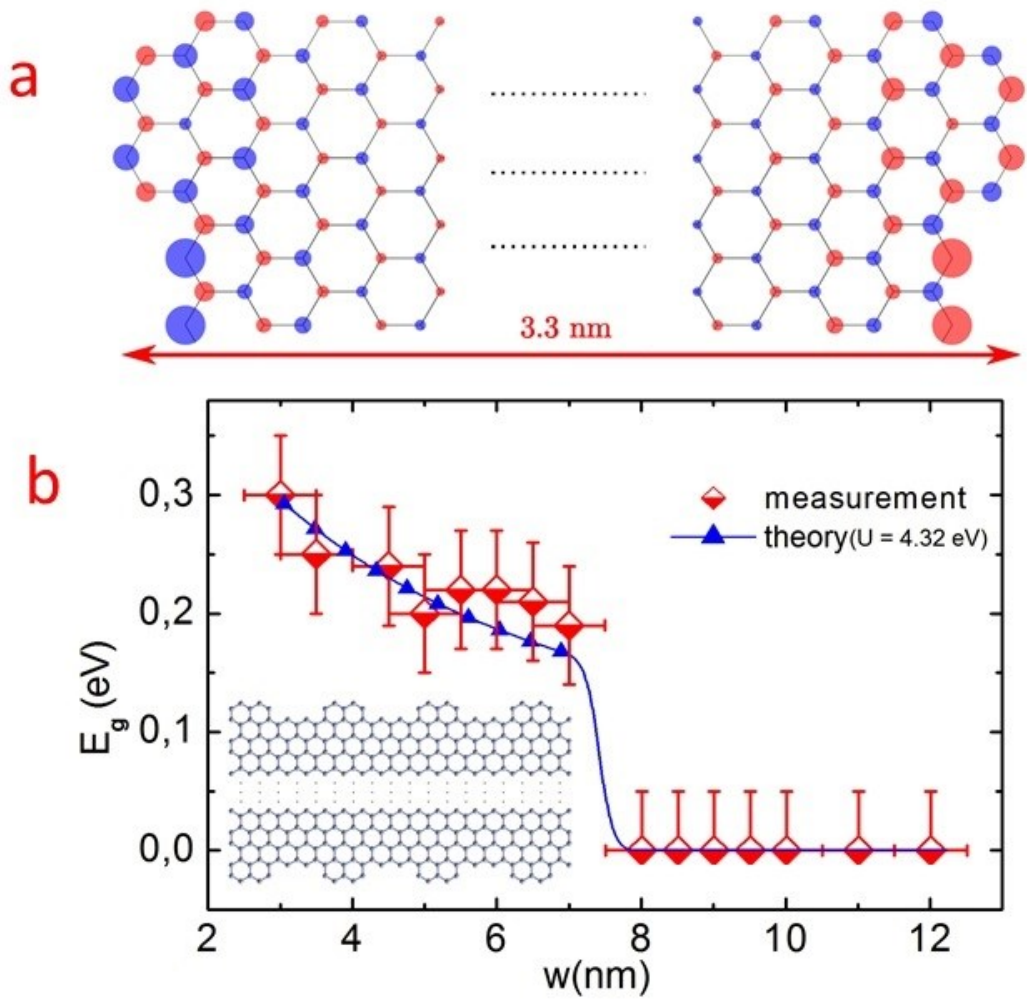

**Extended data: Figure 3**

**Extended data figure 3: The effect of edge irregularities on edge magnetism.** Calculated *spin density distribution in the unit cell of a 3.3 nm wide zigzag ribbon with a high density of atomic scale defects revealing the substantial decrease of the emerging spin polarization (about 1/3 as compared to defect free zigzag edges). The experimental width dependence can be fit for defective ribbon edges by using higher values of the on-site repulsion parameter of U = 4.32 eV.*

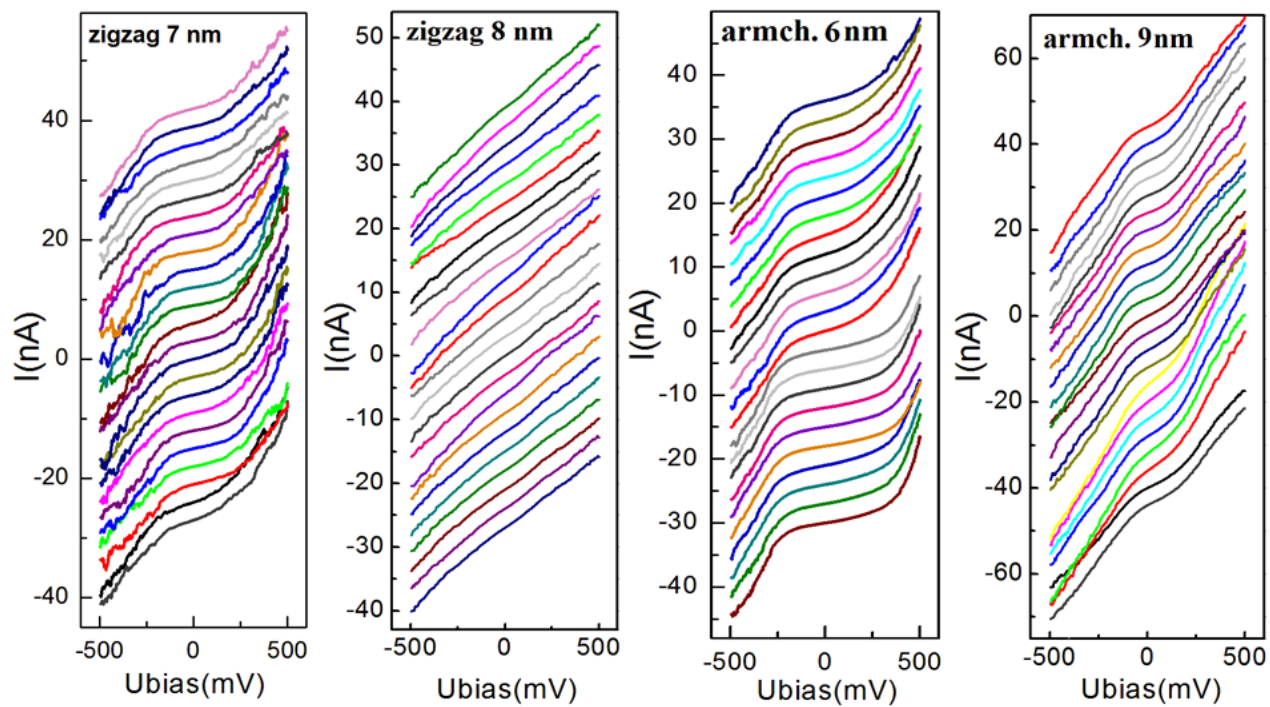

**Extended data: Figure 4**

**Extended data figure 4: Individual tunneling *I-V* spectra.** Tunneling current-voltage characteristics recorded on various ribbons (the spectra have been shifted along the vertical axis for clarity). Each individual spectrum was recorded as the average of 10 voltage sweeps between +/- 500 mV. The metallic (closely linear) or semiconducting (strongly nonlinear) nature of the ribbons is clearly apparent from the individual tunneling *I-V* characteristics.